# Integrating Node Importance and Network Topological Properties for Link Prediction in Complex Network

Zhu Junxi, Dai Fang*, Zhao Fengqun, Guo Wenyan

School of Science, Xi'an University of Technology, Xi'an,710048, China

**Abstract**—Link prediction is one of the most important and challenging tasks in complex network analysis, which aims to predict the likelihood of the existence of missing links based on the known information in the network. As critical topological properties in the network, node degree and clustering coefficient are well-suited for describing the tightness of connection between nodes. The node importance can affect the possibility of link existence to a certain extent. By analyzing the impact of different centrality on links, which concluded that the degree centrality and proximity centrality have the greatest influence on link prediction. So, a link prediction algorithm combines node importance and attribute, called DCCLP, is proposed in this paper. In the training phase of the DCCLP algorithm, the maximized AUC indicator in the training set as the objective, and the optimal parameters are estimated by utilizing the White Shark Optimization algorithm. Then the prediction accuracy of the DCCLP algorithm is evaluated in the test set. By experimenting on twenty-one networks with different scales, and comparing with existing algorithms, the experimental results show that the effectiveness and feasibility of DCCLP algorithm, and further illustrate the importance of the degree centrality of node pairs and proximity centrality of nodes to improve the prediction accuracy of link prediction.

**Key Words**   complex networks; link prediction; proximity centrality; degree centrality; node attribute; clustering coefficient

## 1 INTRODUCTION

Complex systems can be described as complex network that consisting of lots of nodes and edges, then the processing of information in complex systems can be transformed into the mining of information in complex network. Therefore, researching complex networks in a scientific way can help us better recognize and understand the internal structure of complex systems. The research contents of complex network main include community detection[1], link prediction[2], important node ranking[3] and network evolution [4]etc. Link prediction in the network refers to how to predict the possibility of edges existence between nodes which have no connecting edge in the network through known network structure and node attributes. This prediction includes both the prediction of unknown connecting edges and the prediction of future connecting edges[5]. Predicting existing but missing edges is a data mining process, and the prediction of possible future edges is related to the evolution of complex network.

According to the different methods of network feature extraction, the link prediction methods are roughly divided into three categories: network topology structure, maximum likelihood probability and machine learning[6-8]. Although the methods based on maximum likelihood probability and machine learning can obtain better prediction accuracy, their application range is limited due to high computational complexity. The methods based on network topology structure are easier to implement when utilizing network topology to predict the link existence. Existing methods based on network topology can be divided into three categories: local similarity, quasi-similarity, and global similarity.

The link prediction based on local similarity mainly considers the common neighbors' information of nodes in network. The literature[9-11] pay more attention to the degree of common neighbors and the number of common neighbors to predict the network edges. Gao et. al[12] proposed a link prediction algorithm that considers nodes' attribute, which uses the degree and clustering coefficient of common neighbors to estimate

---

*Dai Fang*, corresponding author, E-mail: daifang@xaut.edu.cn

the likelihood of the existence of links between nodes and achieved good prediction accuracy. Yu et. al[13] proposed a link prediction algorithm which combines degree, clustering coefficient, node centrality into link prediction of complex networks. The advantage of the method based on local similarity is low computational complexity. However, due to the limited information used, the prediction accuracy is not ideal.

The link prediction based on global similarity take advantage of more information in the network, such as the Katz method[14], which considered the transfer of similarity of all paths in the network and improved the prediction accuracy of link prediction. Based on the random walk in the network, Brin et. al [15] proposed a random walk method with restart, which improved the prediction accuracy of link prediction. The methods based on global similarity have good prediction accuracy but high computational complexity. In order to improve the prediction accuracy of the local similarity methods and reduce the computational complexity of the global similarity methods, some scholars proposed link prediction methods based on quasi-local similarity, such as the local path method proposed by Zhou et. al[16]. This method considered the contribution of the third-order paths to the node similarity on the basis of the common neighbor method, and has good prediction accuracy. Liu et. al[17] proposed the LRW method based on local random walk of the network. Because only a limited number of steps are used, the computational complexity of this method is low, and it is significantly suitable for link prediction of large-scale networks. Based on the local topology of the network and the strong connection between nodes, and considered the contribution of high-order paths to node similarity, Qian et. al[18] proposed three algorithms: TPSR2, TPSR3 and TPSR4. The experimental results show that these three algorithms have good prediction accuracy, and TPSR3 algorithm performs better.

Inspired by the idea of employing local link information for link prediction, Wu et. al[19] proposed a new similarity method named CCLP, which used more local structure information in the network. Through comparative experiments in networks from various fields, the results show that it is more effective in predicting missing links and has better prediction accuracy. Based on the degree distribution and local information of nodes to estimate the likelihood of links existence between nodes in the network, Shi et. al[20] proposed the CN2D algorithm for link prediction that utilizing network structure information to predict the existence of links. The CN2D algorithm has low computational complexity and high prediction accuracy. Liu et. al[21] based on the initial information contribution of nodes proposed a link prediction algorithm IICN which aimed to solve the problem of ignoring the initial information size of nodes in the information transmission process between nodes. The experimental results demonstrate that, compared with mainstream benchmark methods ,the IICN algorithm has great advantages in effectiveness and robustness.

In this paper, we propose a new link prediction algorithm which considering the degree centrality of node pair and closeness centrality of nodes, called DCCLP, that predict the possibility of link existence from two aspects: node importance and node topological feature. Our DCCLP algorithm makes up the deficiency that TPSR3 algorithm doesn't fully consider the node importance and achieves better prediction accuracy and performance.

## 2 THE TPSR3 ALGORITHM

In 2017, Qian et. al[18] utilized the strong connection of the ego network and the close relationship between nodes to describe the similarity, and proposed a class of TPSR algorithm that combine topological properties and strong ties. The main ideas of these algorithms are using the paths within third-order in the network and the attribute information(such as degree and clustering coefficient)of nodes to describe the similarity between nodes. Through comparison and analysis of experimental results, it is found that TPSR3 algorithm performs best.

For two nodes $x, y \in V$ in the undirected unweighted netwrok $G = (V, E)$, let $L_{x,y}^3$ denote the set of third-order paths of connecting node $x$ and node $y$, then the similarity between nodes $x$ and $y$ is shown in formula (1)[18].

$$s_{xy}^{TPSR3} = \sum_{z \in \Gamma(x) \cap \Gamma(y)} \frac{C_z}{k_z} + \frac{1}{k_z} + \theta |L_{xy}^3| \qquad (1)$$

where, $\Gamma(x), \Gamma(y)$ represent the neighbor node sets of nodes $x$ and $y$ respectively, the parameter $\theta$ is a small positive number, which is used to control the contribution of third-order paths information to node similarity. $|L_{x,y}^3|$ represents the number of elements contained in $|L_{x,y}^3|$, $k_z$ is the degree of node $z$. $C_z$ denote the clustering coefficient of node $z$, and the calculation method is shown in formula (2).

$$C_z = \frac{2t_z}{k_z(k_z-1)} \qquad (2)$$

where $t_z$ represents the number of connected edges between neighbor nodes of node $z$.

According to formula (1), it can be seen that when measures the similarity between nodes in the network, TPSR3 algorithm considered the attributes information of common neighbors of predicted nodes, it also considered the path information between nodes and adjusts the contribution of third-order paths to node similarity by parameter $\theta$.

## 3 THE DCCLP ALGORITHM

Similar to existing link prediction algorithms based on the local network structure, the TPSR3 algorithm did not fully consider the importance of nodes in the network, the importance of nodes has an important effect on the possibility of link's existence between nodes. By analyzing the impact of different centrality on link's existence, literature[22] points out that degree centrality has the most significant effect and the proximity centrality of nodes can also reflect the importance of nodes in the network. The degree centrality of node pairs and proximity centrality of nodes are integrated into TPSR3 algorithm to improve the TPSR3 algorithm in this paper, named DCCLP.

For two nodes $x, y \epsilon V$ in the undirected unweighted network $G(V,E)$, we define the similarity between them as shown in formula (3).

$$s_{xy}^{DCCLP} = \alpha \sum_{z \in \Gamma(x) \cap \Gamma(y)} \left( \frac{C_z}{k_z} + \frac{1}{k_z} + \frac{P_z}{k_z} \right) + \beta D_{xy} + \theta |L_{xy}^3| \quad (3)$$

where $\alpha > 0, \beta > 0$ are adjustable parameters, and $\alpha + \beta = 1$. The parameter $\theta$ is used to control the contribution of the third-order paths to the node similarity, $k_z$ represents the degree of nodes $z$, $D_{xy}$ represents th degree centrality of node pair $(x, y)$, and the calculation method is shown in formula (4).

$$D_{xy} = \frac{k_x k_y}{(|V|-1)^2} \quad (4)$$

where, $|V|$ represents the number of nodes in the network $G$.

In the network $G$, the average shortest path distance from node $z$ to other nodes is expressed as formula (5).

$$d_z = \frac{1}{N} \sum_{\substack{j \in V \\ z \neq j}} d_{zj} \quad (5)$$

Among them, $d_{zj}$ represents the distance of shortest path between node $z$ and node $j$, and the smaller $d_z$, the more important node $z$ is in the network $G$.

The proximity centrality $P_z$ of node $z$ in the network $G$ can be calculated by formula(6).

$$P_z = \frac{1}{d_z} \quad (6)$$

In the following, we describe the detail steps of our DCCLP algorithm according to training phase and testing phase.

**(1) Training Phase**

**Step1** For a given network $G$, 90% of the edges and all the nodes in the network $G$ are randomly selected as the training set, which is satisfied the connectivity of edges.

**Step2** Taking the maximized AUC indicator in the training set as the objective, the adjustable parameters $\theta$ and α are estimated by White Shark intelligent optimization algorithm.

**Step3** Repeating Step1 and Step2 for one hundred independent experiments to calculate the average of AUC, according to the maximum AUC indicator to find the corresponding optimal parameters $\theta^*$ and $\alpha^*$.

**(2) Testing Phase**

Substituting the optimal parameters $\theta^*$ and $\alpha^*$ into formula (3) and selecting 10% of the connected edges in the network $G$ randomly as the test set, selecting the evaluation indicators AUC and Precision to measure the prediction accuracy of the DCCLP algorithm.

## 4 EXPERIMENTAL RESULT AND ANALYSIS

Next, we do two groups of comparative link prediction experiments to verify the effectiveness and feasibility of our DCCLP algorithm.

(1) Comparison of DCCLP algorithm with the algorithms in literature [18]

To verify the effectiveness and prediction accuracy of the DCCLP algorithm, we compare it with the nine algorithms mentioned in the literature [18] in different scale networks. Before the experiment, the dataset is preprocessed, that is, the directed network is converted to the undirected network, and the weight of the edges in weighted network is not considered. Tab.1 and Fig.1(left) show the AUC indicator of ten algorithms. Tab.2 and Fig.1(right) show the Precision indicator of ten algorithms. The black bold values in the Tab.1 and Tab.2 represent the best prediction accuracy among the ten algorithms. The first nine columns of data are from the literature[18]. The full name of AUC indicator is Area Under the Receiver Operating Characteristic Curve, which is quoted from literature[23], and Precision is quoted from literature[24].

According to the above experimental results, our DCCLP algorithm has better prediction accuracy and predictive performance than other nine compared algorithms.

Observed the AUC results in Tab.1, on the nine networks, the DCCLP algorithm obtains the optimal AUC indicator, especially on the Netscience network. Compared with the other nine algorithms, the AUC value is improved about 6%. Observed the Precision results in Tab.2, on nine networks, the DCCLP algorithm obtains the best Precision indicator, especially on FWEW network, the Precision is also improved about 6% compared with the best of the other nine algorithms. By analyzing the advantages of our DCCLP algorithm, there are mainly the following three aspects. Firstly, compared with the local similarity methods, for example: CN, RA, AA, TPSR2, these methods mainly use node attribute information to predict edges existence, our DCCLP algorithm considers the importance of nodes in the network. Secondly, the global similarity method Katz predicts the existence of edges by considering all the paths in the network, our DCCLP algorithm uses the fewer paths information to reduce the computational complexity of algorithm. Thirdly, our DCCLP algorithm considers the degree attribute of predicted node itself when measuring the similarity between nodes. Our DCCLP algorithm performs best in comparison algorithms.

Tab.1 Prediction performance of link prediction methods measured by AUC in a set of real-world networks

| Networks | CCN | TPSR2 | TPSR3 | TPSR4 | CN | AA | RA | LP(0.01) | Katz(0.01) | DCCLP($\theta^*, \alpha^*$) |
|---|---|---|---|---|---|---|---|---|---|---|
| Jazz | 0.9710 | 0.9710 | 0.9220 | 0.9130 | 0.9550 | 0.9620 | 0.9710 | 0.9470 | 0.9420 | **0.9716**(0.0001,1) |
| USAir | 0.9530 | 0.9540 | 0.9260 | 0.9200 | 0.9370 | 0.9480 | 0.9530 | 0.9270 | 0.9240 | **0.9703**(0.0013,0.9798) |
| C.elegans | **0.8980** | 0.8640 | 0.8710 | 0.8630 | 0.8430 | 0.8600 | 0.8630 | 0.8590 | 0.8560 | 0.8602(0.0422,0.9242) |
| FWFW | 0.7580 | 0.6250 | 0.8080 | 0.7860 | 0.6110 | 0.6130 | 0.6160 | 0.6720 | 0.6790 | **0.8143**(0.0943,0.0290) |
| FWFD | 0.7570 | 0.6200 | 0.8080 | 0.7860 | 0.6100 | 0.6100 | 0.6130 | 0.6720 | 0.6800 | **0.8168**(0.0819,0.0478) |
| FWEW | 0.8120 | 0.7130 | 0.8410 | 0.8320 | 0.6920 | 0.6990 | 0.7070 | 0.7360 | 0.7410 | **0.8505**(0.0713,0.0042) |
| FWMW | 0.7930 | 0.7140 | 0.8180 | 0.8040 | 0.7020 | 0.7070 | 0.7110 | 0.7390 | 0.7400 | **0.8249**(0.0829,0.0571) |
| PB | 0.9420 | 0.9230 | 0.9320 | 0.9270 | 0.9190 | 0.9210 | 0.9230 | 0.9300 | 0.9240 | **0.9428**(0.0366,0.7578) |
| Netscience | 0.9400 | 0.9350 | 0.9400 | 0.9400 | 0.9360 | 0.9360 | 0.9350 | 0.9400 | 0.9400 | **0.9989**(0.0822,0.0160) |

Tab. 2 Prediction performance of link prediction methods measured by Precision in a set of real-world networks

| Networks | CCN | TPSR2 | TPSR3 | TPSR4 | CN | AA | RA | LP(0.01) | Katz(0.01) | DCCLP($\theta^*, \alpha^*$) |
|---|---|---|---|---|---|---|---|---|---|---|
| Jazz | 0.8240 | 0.8380 | 0.6500 | 0.6000 | 0.8190 | 0.8380 | 0.8240 | 0.7840 | 0.8090 | **0.8385**(0.0001,1) |
| USAir | 0.6270 | 0.6310 | 0.5710 | 0.5610 | 0.5910 | 0.6070 | 0.6270 | 0.5860 | 0.5900 | **0.6374**(0.0013,0.9798) |
| C.elegans | 0.1430 | 0.1330 | **0.1590** | 0.1520 | 0.1320 | 0.1340 | 0.1330 | 0.1360 | 0.1360 | 0.1587(0.0422,0.9242) |
| FWFW | 0.1600 | 0.0880 | 0.3420 | 0.2900 | 0.0900 | 0.0900 | 0.0860 | 0.1300 | 0.0980 | **0.3868**(0.0943,0.0290) |
| FWFD | 0.1680 | 0.0910 | 0.3510 | 0.2990 | 0.0900 | 0.0890 | 0.0870 | 0.1310 | 0.0950 | **0.3955**(0.0819,0.0478) |
| FWEW | 0.2600 | 0.1700 | 0.3200 | 0.3060 | 0.1470 | 0.1570 | 0.1650 | 0.1850 | 0.1580 | **0.3768**(0.0713,0.0042) |
| FWMW | 0.2270 | 0.1470 | 0.3340 | 0.3010 | 0.1390 | 0.1440 | 0.1450 | 0.1740 | 0.1470 | **0.3818**(0.0829,0.0571) |
| PB | 0.2980 | 0.2470 | 0.5030 | 0.4830 | 0.4050 | 0.3610 | 0.2400 | 0.4430 | 0.4130 | **0.5075**(0.0366,0.7578) |
| Netscience | 0.9700 | **0.9720** | 0.9710 | 0.9710 | 0.8160 | 0.9660 | 0.9640 | 0.8100 | 0.8100 | 0.8409(0.0822,0.0160) |

The test sets contain $q = 0.1$ fraction of the edges in the complete networks and the presented results are the average of 100 independent runs. For the DCCLP algorithm, the values in parentheses represent the optimal value of adjustable parameters $\theta^*$ and $\alpha^*$, the parameter in LP method and Katz method are set 0.01.

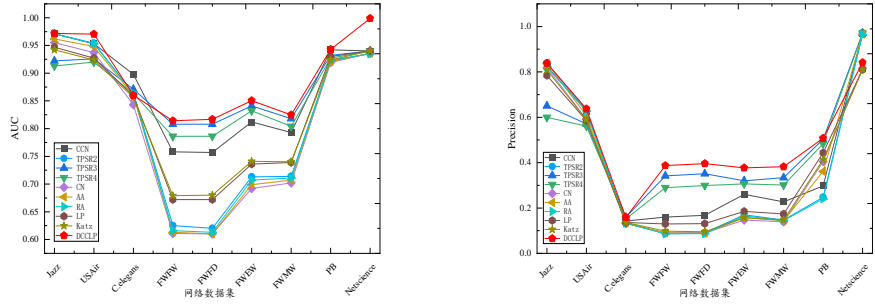

Fig. 1 The average AUC and Precision of the DCCLP algorithm and the corresponding link prediction algorithms based on similarity. For almost all networks (expect C.elegans and NetScience), the DCCLP algorithm surpass the all comparison algorithms.

(2) Comparison of the DCCLP algorithm with existing algorithms

To further verify the effectiveness and predictive performance of the DCCLP algorithm, we compared with the CCNC[13], CCLP[19], CN2D[20] algorithms which based on the network topology, and they ignore the influence of node degree centrality on link existence in the network. The twenty-one networks of different scales are collected for comparative experiments. These networks involve social network, biological network, citation network, cooperative network, aviation network and other network. The topological properties of the networks are shown in Tab. 3.

The DCCLP algorithm proposed in this paper is used for link prediction in twenty-one networks shown in Tab.3, and compared with the CCNC, CCLP and CN2D algorithms, the prediction results are shown in Tab.4, Tab.5, and Fig.2. The Tab.4 and Fig.2(left) show the AUC indicator of four algorithms, and Tab. 5 and Fig.2(right) show the Precision indicator of four algorithms.

Tab. 3 Basic topological properties of twenty-one networks used in the experiments

| Networks | $|V|$ | $|E|$ | $\langle k \rangle$ | $\langle C \rangle$ | $D$ | $H$ | $\langle d \rangle$ |
|---|---|---|---|---|---|---|---|
| Polbook | 105 | 441 | 8.400 | 0.4875 | 0.0808 | 1.4207 | 3.0788 |
| Dolphin | 62 | 159 | 5.129 | 0.3852 | 0.0841 | 1.3243 | 3.1089 |
| karate | 34 | 78 | 4.5882 | 0.6001 | 0.1390 | 1.6933 | 2.4082 |
| FWEW | 69 | 880 | 25.5072 | 0.5521 | 0.3751 | 1.2746 | 1.636 |
| FWFW | 128 | 2075 | 32.4219 | 0.3346 | 0.2553 | 1.2370 | 1.7763 |
| FWMW | 97 | 1446 | 29.8144 | 0.4683 | 0.3106 | 1.2656 | 1.6929 |
| football | 115 | 613 | 10.6609 | 0.4032 | 0.0935 | 1.0069 | 2.5082 |
| Grassland | 75 | 114 | 3.0400 | 0.8198 | 0.0411 | 2.7499 | 3.1996 |
| Trainbombing | 64 | 243 | 7.5938 | 0.7473 | 0.1205 | 1.6588 | 2.691 |
| C.elegans | 297 | 2148 | 14.4646 | 0.3429 | 0.0489 | 1.8008 | 2.4553 |
| USAir | 332 | 2126 | 12.8072 | 0.7909 | 0.0387 | 3.4639 | 2.7381 |
| Infectious | 410 | 2765 | 13.4878 | 0.4802 | 0.0330 | 1.3876 | 3.6309 |
| FWFD | 128 | 2106 | 32.9063 | 0.3346 | 0.2591 | 1.2307 | 1.7724 |
| Metabolic | 453 | 2025 | 8.9404 | 0.6597 | 0.0198 | 4.485 | 2.6638 |
| Jazz | 198 | 2742 | 27.6970 | 0.6427 | 0.1406 | 1.3951 | 2.235 |
| US Roads | 49 | 107 | 4.3673 | 0.5171 | 0.0910 | 1.1299 | 4.1633 |
| PB | 1222 | 16714 | 27.3552 | 0.4307 | 0.0224 | 2.9707 | 2.7375 |
| Netscience | 1589 | 2742 | 3.4512 | 0.8310 | 0.0022 | 2.0105 | 0.3514 |
| Email | 1133 | 5451 | 9.6222 | 0.3535 | 0.0085 | 1.9421 | 3.606 |
| Bio-DM-LC | 658 | 1129 | 3.4316 | 0.5166 | 0.0052 | 3.1149 | 3.5637 |
| Bio-CE-GT | 924 | 3239 | 7.0108 | 0.6820 | 0.0076 | 4.1392 | 3.3724 |

where $|V|$ is the number of nodes in the network; $|E|$ is the number of edges in the network; $\langle k \rangle$ is the average degree of the network; $\langle C \rangle$ is the average clustering coefficient of the network; $D$ is the network density; $H$ is the network degree heterogeneity; $\langle d \rangle$ is the average shortest distance of the network.

Tab.4 and Tab. 5 respectively show the four algorithms' results of AUC and Precision indicators in twenty-one networks. The values in parentheses in Tab.4 and Tab.5 represent the optimal parameters $\theta^*$ and $\alpha^*$. The experimental data of the first three algorithms are cited in the corresponding literature, indicating that this data is from the literature, and the rest are the results of this paper.

Tab. 4 The AUC indicator for link prediction of four algorithms in twenty-one networks

| Networks | CN2D | CCLP | CCNC | DCCLP($\theta^*, \alpha^*$) |
| --- | --- | --- | --- | --- |
| Polbook | 0.8791 | 0.8908 | **0.9240**[13] | 0.8876(0.0826,0.7284) |
| Dolphin | 0.7722 | 0.8020[19] | **0.8360**[13] | 0.7981(0.0883,0.0735) |
| karate | 0.6441 | 0.6960 | 0.7332 | **0.7925**(0.0960,0.0784) |
| FWEW | 0.6750 | 0.7026 | 0.7216 | **0.8505**(0.0713,0.0042) |
| FWFW | 0.6042 | 0.6362[19] | 0.6470[13] | **0.8143**(0.0943,0.290) |
| FWMW | 0.7079 | 0.7229 | 0.7267 | **0.8249**(0.0829,0.0571) |
| football | **0.8535** | 0.8397 | 0.8420 | 0.8472(0.0834,0.002) |
| Grassland | 0.8236 | 0.7900[19] | **0.8987** | 0.8655(0.0891,0.0606) |
| Trainbombing | 0.9283 | 0.9317 | **0.9424** | 0.9328(0.0296,0.9077) |
| C.elegans | 0.8613[20] | 0.8658 | **0.8721** | 0.8602(0.0422,0.9242) |
| USAir | 0.9695[20] | 0.9576 | 0.9620[13] | **0.9703**(0.0013,0.9798) |
| Infectious | 0.9393 | 0.9399 | 0.9447 | **0.9579**(0.0156,0.6831) |
| FWFD | 0.6003 | 0.6308 | 0.6318 | **0.8168**(0.0819,0.0478) |
| metabolic | 0.9039 | 0.9507 | **0.9592** | 0.9542(0.0034,1.0000) |
| Jazz | 0.9685[20] | 0.9600[19] | **0.9740**[13] | 0.9716(0.0001,1.000) |
| USRoads | **0.9007** | 0.8647 | 0.8829 | 0.8182(0.0892,0.0566) |
| Email | 0.8546 | 0.8570[19] | 0.8578 | **0.9168**(0.0504,0.5232) |
| bio-DM-LC | 0.6701 | 0.6498 | 0.6707 | **0.9684**(0.0411,0.0007) |
| bio-CE-GT | 0.9341 | 0.9403 | 0.9547 | **0.9717**(0.0168,0.7671) |
| PB | **0.9599**[20] | 0.9266[19] | 0.9360[13] | 0.9397(0.0366,0.7578) |
| Netscience | 0.9981[20] | 0.9480 | 0.9920 | **0.9989**(0.0822,0.0160) |

Tab. 5 The Precision indicator for link prediction of four algorithms in twenty-one networks

| Networks | CN2D | CCLP | CCNC | DCCLP($\theta^*, \alpha^*$) |
| --- | --- | --- | --- | --- |
| Polbook | 0.1145 | 0.1426 | **0.1485** | 0.1257(0.0826,0.7284) |
| Dolphin | **0.0646** | 0.0528 | 0.0576 | 0.0551(0.0883,0.0735) |
| karate | 0.0307 | 0.0385 | 0.0424 | **0.0489**(0.0960,0.0784) |
| FWEW | 0.1423 | 0.1671 | 0.1978 | **0.3768**(0.0713,0.0042) |
| FWFW | 0.0853 | 0.0970[19] | 0.1068 | **0.3868**(0.0943,0.290) |
| FWMW | 0.1320 | 0.144 | 0.1735 | **0.3818**(0.0829,0.0571) |
| football | **0.3112** | 0.2619 | 0.2738 | 0.2477(0.0834,0.002) |
| Grassland | 0.0441 | 0.0565 | **0.0698** | 0.0465(0.0891,0.0606) |
| Trainbombing | 0.1791 | 0.1912 | **0.2083** | 0.1777(0.0296,0.9077) |
| C.elegans | 0.1213 | 0.1356 | 0.1316 | **0.1587**(0.0422,0.9242) |
| USAir | 0.6046 | 0.6166 | **0.6503** | 0.6374(0.0013,0.9798) |
| Infectious | 0.3803 | 0.3592 | **0.5083** | 0.2720(0.0156,0.6831) |
| FWFD | 0.0862 | 0.0967 | 0.1109 | **0.3955**(0.0819,0.0478) |
| metabolic | 0.1918 | 0.2467 | **0.3188** | 0.2929(0.0034,1.0000) |
| Jazz | 0.8243 | **0.8590**[19] | 0.8297 | 0.8385(0.0001,1.000) |
| USRoads | **0.0800** | 0.0603 | 0.0657 | 0.0238(0.0892,0.0566) |
| Email | 0.2953 | **0.3080**[19] | 0.2878 | 0.1850(0.0504,0.5232) |
| bio-DM-LC | 0.0043 | 0.0456 | 0.0207 | **0.4751**(0.0411,0.0007) |
| bio-CE-GT | 0.1076 | 0.1911 | 0.2228 | **0.2961**(0.0168,0.7671) |
| PB | 0.4288 | 0.4040[19] | 0.2732 | **0.5075**(0.0366,0.7578) |
| Netscience | 0.8464 | **0.8982** | 0.6241 | 0.8409(0.0822,0.0160) |

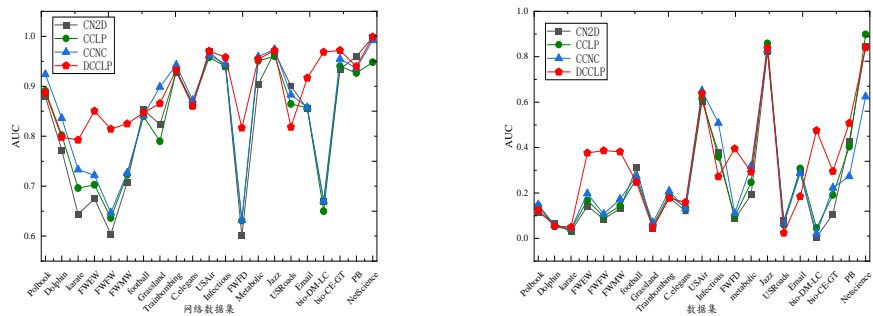

Fig. 2 The average AUC and Precision of the DCCLP algorithm and the comparative link prediction algorithms based on the network topology.

From the above experimental results, among the twenty-one networks shown in Tab. 3, there are eleven networks with the best AUC indicator, and nine networks with best Precision indicator. This further show that the superiority of the DCCLP algorithm. Compared with the other three algorithms, our DCCLP algorithm performs better in biological networks, cooperative networks and aviation networks by introducing the node importance. The prediction accuracy in social networks is not as good as CCNC algorithm.

The experimental results show that, compared with CCNC, CCLP and CN2D algorithms, the DCCLP algorithm has better prediction accuracy and better prediction effect. At the same time, the results also show that the clustering coefficient, degree, proximity centrality of common neighbor nodes and the degree centrality of predicted nodes can promote possibility of connecting edges between nodes and can well reflect the difference degree of the contribution ability of each neighbor node.

## 5 CONCLUSIONS

The degree centrality of predicted nodes and the proximity centrality of nodes are integrated in TPSR3 algorithm, and the link prediction algorithm DCCLP is proposed in this paper. The DCCLP algorithm predicts the existence of unknown edges in the network from the perspectives of network structure similarity and node importance. Through the two groups comparison experiments, the prediction accuracy AUC and Precision of the DCCLP algorithm outperformance in all comparison algorithms', which verifies the effectiveness and feasibility of the DCCLP algorithm, and further illustrates the importance of degree centrality and proximity centrality to improve the prediction accuracy for link prediction.

In this paper, we mainly consider the influence of node importance on link prediction in undirected unweighted networks, but the research on node importance in directed networks and weighted networks still needs to continue being explored. In the following research, we will generalize our algorithm to directed networks and weighted networks.

## ACKNOWLEDGMENT

This work was supported by the National Natural Science Foundation of China [ grant number 61976176].